\documentclass[%
reprint,
superscriptaddress,
amsmath,amssymb,
aps,
]{revtex4-2}
\usepackage{graphicx}
\usepackage{dcolumn}
\usepackage{bm}
\usepackage{xcolor}
\usepackage{siunitx}

\begin{document}


\title{Polaritonic Ultrastrong Coupling: Quantum Entanglement in Ground State}

\author{Qingtian Miao}
\email{qm8@tamu.edu}
  \affiliation{Department of Physics and Astronomy, Texas A\&M University, Texas 77843, USA}
\author{G.S. Agarwal}
 \affiliation{Institute for Quantum Science and Engineering, Department of Biological and Agricultural Engineering, Department of Physics and Astronomy, Texas A\&M University, College Station, Texas 77843, USA}

\date{\today}

\begin{abstract}
The ultrastrong coupling between the elementary excitations of matter and microcavity modes is studied in a fully analytical quantum-mechanical theoretical framework. The elementary excitation could be phonons, excitons, plasmons, etc. From the diagonalization of the Hamiltonian, we obtain the ground state of the polariton Hamiltonian. The ground state belongs to the Gaussian class. Using the Gaussian property we calculate the quantum entanglement in the ground state. We use two different measures for quantum entanglement --- entanglement entropy and the logarithmic negativity parameter and obtain rather simple analytical expressions for the entanglement measures. Our findings show that the amount of quantum entanglement in the ground state is quite significant in the ultrastrong coupling regime. It can be obtained from the measurement of the polariton frequencies.
\end{abstract}

\maketitle

\section{Introduction}

The behavior of matter interacting with cavity fields is known to produce distinct characteristics from that in free space \cite{Haroche2006,Walther2002,GSAJMO}. This is because the interaction parameter $g$ can have values in the weak coupling range, or strong or even ultrastrong range. While the strong coupling regime has been thoroughly investigated, the ultrastrong coupling is attracting intense attention now \cite{Solano2019, Norireview}. In the case of a qubit interacting with a cavity, the ultrastrong coupling $g/\omega$ say about $0.5$ or more is yet to be realized. In such cases, parametric interactions have been suggested to reach the ultrastrong coupling regime \cite{Clerk2018, Nori2018, Ricardo2021}. The situation is different for a large collective system interacting with cavity fields as here one can take advantage of the effective enhancement of the coupling by $\sqrt N$ factor \cite{GSA1984}. Such an enhancement factor has felicitated the observation of strong coupling even when $g$ was in the weak coupling range \cite{Nakamura2014,Hu2017,Liu2015}. Several recent works have successfully observed ultrastrong coupling regime as the density of the elementary excitations was in the range of solid state densities. Some of these recent experiments are for --- phonon polaritons in cavities containing hexagonal boron nitride (hBN) \cite{phonon2021}; plasmon polaritons \cite{Ho2018,plasmon2020}; excitonic systems \cite{KenaCohen2013,exciton2014,exciton2018,Held2018,Gao2018}; 2D electron gas \cite{gas2012}. The possibility of novel photochemistry in cavities under ultrastrong coupling has been investigated \cite{Mukamel2016}. The cavity-free ultrastrong coupling was also realized using meta materials \cite{Cavityfree2018,Cavityfree2021}. A typical experimental observation consists of the observation of the transmission spectra and the polaritonic splitting of the spectra --- the splitting is of the order of $g$ which is about $30-70\%$ of the bare frequency of the cavity which is supposed to be on resonance with the frequency of the elementary excitation in the solid state material. All this can be explained in classical calculations of transmission from a layered medium \cite{Yeh1988}. 

At a quantum level the elementary excitations are coupled to the radiation field modes and the Hamiltonian is diagonalized to find the new quasiparticles, i.e., polaritons \cite{Hopfield,Ciuti2005,Birman1991}. The transmission spectrum basically probes the separation between the frequencies of two polaritons though it has to be borne in mind that the observed separation would also depend on the reflection and transmission properties of the mirrors forming the cavity. This latter aspect has not been fully addressed. In this paper, we study the quantum entanglement which is present in the ground state of the Hopfield Hamiltonian. We derive the ground state which belongs to the class of Gaussian states. We then use two different entanglement measures --- entanglement entropy \cite{Horodecki2009} and the logarithmic negativity parameter \cite{Vidal2001}. The quantum entanglement increases as $g/\omega$ increases. There is no quantum entanglement in the limits of weak and strong couplings as $g/\omega\ll1$. The magnitude of entanglement is given in terms of the polariton frequencies and thus experimental study of the transmission spectra can be used to assess the amount of entanglement.

\section{Polaritons in Ultrastrong Coupling}

In the following, we will consider a microcavity fully filled with hBN where the ultrastrong phonon-photon coupling was demonstrated \cite{phonon2021}. And our theory can also be applied to other systems like exciton-polaritons and plasmon-polaritons in the ultrastrong coupling regime. The transmission spectrum can be measured at normal incidence, where a polaritonic splitting shows the light couples with the in-plane transverse optical (TO) phonon of hBN, which takes place perpendicular to the direction of propagation. Thus we consider ultrastrong coupling of two harmonic oscillators: the normal incidence Fabry\--{}Pérot microcavity mode with energy $\omega_{\rm c }$, and the TO phonon mode with energy $\omega_{0}$ (we set $\hbar=1$). The behavior of the system can be described by the Hopfield Hamiltonian, which reads \cite{Liberato2014}
\begin{equation}
	\begin{aligned}
		\hat{H}&=\omega_{\rm c}\left(\frac{1}{2}+\hat{a}^{\dagger} \hat{a}\right)+\omega_{0}\left(\frac{1}{2}+\hat{b}^{\dagger} \hat{b}\right)\\
		&\qquad +\Omega\left(\hat{a}^{\dagger}+\hat{a}\right)\left(\hat{b}^{\dagger}+\hat{b}\right)+G\left(\hat{a}^{\dagger}+\hat{a}\right)^2,	
		\label{equ:ha}
	\end{aligned}
\end{equation}
where $\hat{a}$ and $\hat{b}$ are the microcavity and TO phonon annihilation operators, respectively, and $\Omega$ is the vacuum Rabi energy. The third term in Eq. (\ref{equ:ha}) contains the anti-resonant or counter-rotating part which cannot be omitted in the ultrastrong coupling regime. The fourth term, which is called the diamagnetic term, originates from the quadratic electromagnetic vector potential term $\mathbf A^2$ of the light-matter minimal coupling. Any quasi-resonant transition gives a contribution to the coefficient $G$, but in this paper, we omit the effect of other transitions, then $G=\Omega^2/\omega_{0}$ \cite{Norireview}. The vacuum Rabi energy for phonon polaritons being considered can be written as \cite{Hopfield} 
\begin{equation}
	\Omega=g\sqrt{\frac{\omega_0}{\omega_{\rm c}}},
\end{equation}
where $g$ is called coupling strength in this article.

The phonon-cavity polaritonic transition energies $\omega_{\pm}$ can be obtained by two-mode Hamiltonian diagonalization,  
\begin{equation}
	\begin{aligned}
		\omega^2_+=\omega_{0}^2+2\Omega\sqrt{\omega_{\rm c}\omega_{0}}\tan\frac{\theta}{2},\\
		\omega^2_-=\omega_{0}^2-2\Omega\sqrt{\omega_{\rm c}\omega_{0}}\cot\frac{\theta}{2},
	\end{aligned}
	\label{equ:en}
\end{equation} 
where $\theta=\arctan\left(\frac{\omega_{\rm c}\left(4G+\omega_{\rm c}\right)-\omega_{0}^2}{4\Omega\sqrt{\omega_{\rm c}\omega_{0}}}\right)+\frac{\pi}{2}$, $(0<\theta<\pi)$, and $\omega_+>\omega_-$. The polariton energies $\omega_\pm$ coincide with the classical bulk dielectric dispersion law \cite{Quattropani1986},
\begin{equation}
	\epsilon(\omega)=\frac{c^2k^2}{\omega^2}=\epsilon_\infty+\frac{4\epsilon_\infty g^2}{\omega_0^2-\omega^2},
	\label{equ:die}
\end{equation} 
where $\epsilon(\omega)$ is the dielectric function of hBN, $\mathbf k$ is the wave vector, and $\omega_{\rm c}=kc/\sqrt{\epsilon_{\infty}}$.

The diagonalized Hamiltonian of the system can be written in a simple form,
\begin{equation}
	H=\omega_-\left(\frac12+\hat A_-^\dagger \hat A_-\right)+\omega_+\left(\frac12+\hat A_+^\dagger \hat A_+\right),
\end{equation} 
where the phonon-cavity polariton normal modes, i.e., the new quasiparticle annihilation operators can be introduced as
\begin{equation}
	\begin{aligned}
		\hat A_-&=-\frac12\cos\frac\theta2\left(\mu+\frac1\mu\right) \hat a+\frac12\sin\frac\theta2\left(\nu+\frac1\nu\right) \hat b\\
		&\qquad-\frac12\cos\frac\theta2\left(\mu-\frac1\mu\right) \hat a^\dagger-\frac12\sin\frac\theta2\left(\nu-\frac1\nu\right) \hat b^\dagger,
	\end{aligned}
\end{equation} 
\begin{equation}
	\begin{aligned}
		\hat A_+&=\frac12\sin\frac\theta2\left(\nu+\frac1\nu\right) \hat a+\frac12\cos\frac\theta2\left(\mu+\frac1\mu\right) \hat b\\
		&\qquad+\frac12\sin\frac\theta2\left(\nu-\frac1\nu\right) \hat a^\dagger-\frac12\cos\frac\theta2\left(\mu-\frac1\mu\right) \hat b^\dagger,
	\end{aligned}
\end{equation} 
with defined variables $\mu=\sqrt{\omega_-/\omega_{\rm c}}$ and $\nu=\sqrt{\omega_+/\omega_{\rm c}}$. To obtain this form of canonical transformation, we also use the simple relation $\omega_-\omega_+=\omega_{\rm c}\omega_{0}$ which can be derived from Eq. (\ref{equ:die}). And we can also write $\hat A_{\pm}$ in terms of quadratures,
\begin{equation}
	\sqrt2 \hat A_-=-\cos\frac{\theta}{2}\left(\mu x_1+\frac1\mu\frac{\partial}{\partial x_1}\right)+\sin\frac{\theta}{2}\left(\frac1\nu x_2+\nu\frac{\partial}{\partial x_2}\right),
\end{equation} 
\begin{equation}
	\sqrt2 \hat A_+=\sin\frac{\theta}{2}\left(\nu x_1+\frac1\nu\frac{\partial}{\partial x_1}\right)+\cos\frac\theta2\left(\frac1\mu x_2+\mu\frac{\partial}{\partial x_2}\right),
\end{equation} 
where the Hermitian quadrature operators in the system are defined as $\hat x_1=\frac{\hat a+\hat a^\dagger}{\sqrt{2}}$, $\hat p_1=\frac{\hat a-\hat a^\dagger}{\sqrt 2i}$, $\hat x_2=\frac{\hat b+\hat b^\dagger}{\sqrt{2}}$, $\hat p_2=\frac{\hat b-\hat b^\dagger}{\sqrt 2i}$.

\section{Ground State}

\begin{figure}[b]
	\includegraphics[width=8.6cm]{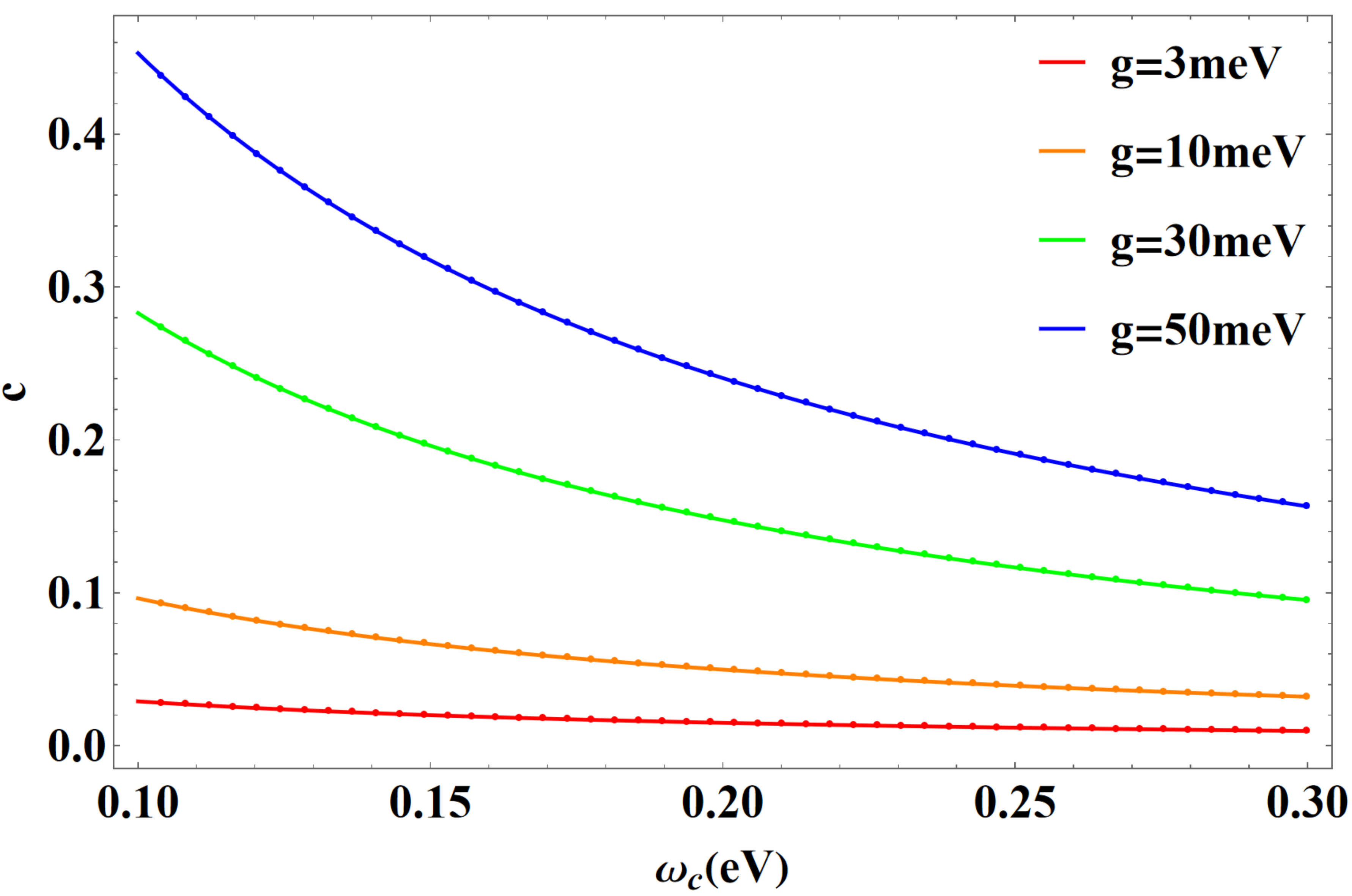}
	\caption{Coefficient $c$ in the Gaussian wave function plotted against the bare cavity energy $\omega_{\rm c}$ with $\omega_{0}=169.1$ meV and different values of $g$.}
	\label{fig:c}
\end{figure}

Now we consider the ground Fock state $|0_{\hat A_-},0_{\hat A_+}\rangle$ of the two-mode system with energy $\omega_{0,0}=\frac{\omega_-+\omega_+}{2}$. From Eq. (\ref{equ:die}), $(\omega_++\omega_-)^2=(\omega_{\rm c}+\omega_{0})^2+4g^2$, thus $\omega_{0,0}>\frac{\omega_{\rm c}+\omega_{0}}{2}$. The ground state possesses virtual photons because our Hamiltonian does not conserve the number of photons. Since it is the vacuum of polariton excitations, $\hat A_-|0_{\hat A_-},0_{\hat A_+}\rangle=0$, $\hat A_+|0_{\hat A_-},0_{\hat A_+}\rangle=0$, and the ground state wave function $\Psi_{0,0}(x_1,x_2)=\langle x_1,x_2|0_{\hat A_-},0_{\hat A_+}\rangle$, we can obtain first order differential equations
\begin{equation}
	\begin{aligned}
		\left[-\cos\frac{\theta}{2}\left(\mu x_1+\frac1\mu\frac{\partial}{\partial x_1}\right)+\sin\frac{\theta}{2}\left(\frac1\nu x_2  +\nu\frac{\partial}{\partial x_2}\right)\right]&\\
		&\!\!\!\!\!\!\!\!\!\!\!\!\!\!\!\!\!\!\!\!\!\!\!\!\!\!\!\Psi_{0,0}(x_1,x_2)=0,
	\end{aligned}
\end{equation} 
\begin{equation}
	\begin{aligned}
	\left[\sin\frac{\theta}{2}\left(\nu x_1+\frac1\nu\frac{\partial}{\partial x_1}\right)+\cos\frac\theta2\left(\frac1\mu x_2+\mu\frac{\partial}{\partial x_2}\right)\right]&\\
	&\!\!\!\!\!\!\!\!\!\!\!\!\!\!\!\!\!\!\!\!\!\!\!\Psi_{0,0}(x_1,x_2)=0.
	\end{aligned}
\end{equation} 
Thus the normalized two-mode Gaussian wave function in the ground state in quadrature space can be calculated by solving those equations,
\begin{equation}
	\Psi_{0,0}(x_1,x_2)=\frac{1}{\sqrt{\pi}}\exp\left(-\frac12\left(ax_1^2+bx_2^2+2c x_1x_2\right)\right),
\end{equation} 
with real coefficients
\begin{equation}
	a=\mu^2\cos^2\frac{\theta}{2}+\nu^2\sin^2\frac{\theta}{2},\quad b=\frac{1}{\mu^2}\cos^2\frac{\theta}{2}+\frac{1}{\nu^2}\sin^2\frac{\theta}{2},
	\label{equ:ab}
\end{equation}
and
\begin{equation}
	c=\frac{\omega_+-\omega_-}{2\sqrt{\omega_{\rm c}\omega_{0}}}\sin\theta=\frac{2\Omega}{\omega_-+\omega_+},
	\label{equ:c}
\end{equation}
where we use the simple relation $\omega_+^2-\omega_-^2=4\Omega\sqrt{\omega_{\rm c}\omega_{0}}/\sin\theta$ which can be derived from Eq. (\ref{equ:en}). The coefficient $c$ directly affects the entanglement, and we plot it as a function of the bare cavity energy $\omega_{\rm c}$ in Fig. \ref{fig:c} for the phonon frequency $\omega_{0}=169.1$ meV of hBN and different coupling strength $g$. Since $c\neq0$, two-mode wavefunction $\Psi_{0,0}(x_1,x_2)\neq f_1(x_1)f_2(x_2)$, where $f_1[f_2]$ is a function of $x_1[x_2]$ alone. Thus we find the ground state of our system is inseparable or entangled. Clearly, to have significant entanglement, $g$ cannnot be too small. For weak coupling, i.e., $g\rightarrow0$, the polariton normal modes go back to cavity mode and phonon mode, and we can expect that the amount of entanglement decreases to $0$. 

We can also calculate the first excited states in quadrature space,
\begin{equation}
	\begin{aligned}
		\Psi_{1,0}(x_1,x_2)&=A_-^\dagger\Psi_{0,0}(x_1,x_2)\\
		&=\sqrt2\left(-\mu\cos\frac{\theta}{2}x_1+\frac{1}{\nu}\sin\frac{\theta}{2}x_2\right)\Psi_{0,0}(x_1,x_2),
	\end{aligned}
\end{equation}
\begin{equation}
	\begin{aligned}
		\Psi_{0,1}(x_1,x_2)&=A_+^\dagger\Psi_{0,0}(x_1,x_2)\\
		&=\sqrt2\left(\nu\sin\frac\theta2 x_1+\frac{1}{\mu}\cos\frac\theta2x_2\right)\Psi_{0,0}(x_1,x_2),
	\end{aligned}
\end{equation}
with one polariton excitation. $\Psi_{1,0}(x_1,x_2)$ is the wavefunction for the lower polariton, while $\Psi_{0,1}(x_1,x_2)$ represents the upper polariton.

\section{Quantitative Measures of Entanglement}

\subsection{Entanglement Entropy}

For a general pure bipartite quantum state, the Von Neumann entropy of either of the reduced density matrix of a subsystem serves as entanglement entropy to measure the degree of quantum entanglement, and it can be proved that they have the same value. For our system, the ground state density operator is $\hat\rho=|0_{\hat A_-},0_{\hat A_+}\rangle\langle0_{\hat A_-},0_{\hat A_+}|$. Since it is a pure state, the entropy of the bipartite state $S(\hat a,\hat b)=0$, and the entropies of the reduced density matrices are the same, $S(\hat a)=S(\hat b)$ \cite{Rendell2005}. The ground state is Gaussian and hence the state of each mode will be a mixed Gaussian state which has a Gaussian Wigner function in quadrature space,
\begin{equation}
	W(x,p)=\frac{1}{2\pi\sqrt{\alpha\beta-\gamma^2}}\exp\left(-\frac12\frac{\alpha x^2+\beta p^2-2\gamma x p}{\alpha\beta-\gamma^2}\right),
\end{equation}
where the variance $\alpha=\langle\hat p^2\rangle$, $\beta=\langle\hat x^2\rangle$, $\gamma=\frac12\langle\hat x\hat p+\hat p\hat x\rangle$, with no displacements, i.e., $\langle \hat x\rangle=\langle\hat p\rangle=0$. Then the expression for entropy of a reduced density operator of either mode can be written as \cite{Agarwal1971,Rendell2005}
\begin{equation}
	S=k_B\left[\left(\sigma+1\right)\ln\left(\sigma+1\right)-\sigma\ln\sigma\right],
\end{equation}
where $k_B$ is the Boltzmann constant, and
\begin{equation}
	\sigma=\left(\alpha\beta-\gamma^2\right)^{1/2}-\frac12.
\end{equation}

\begin{figure}[b]
	\includegraphics[width=8.6cm]{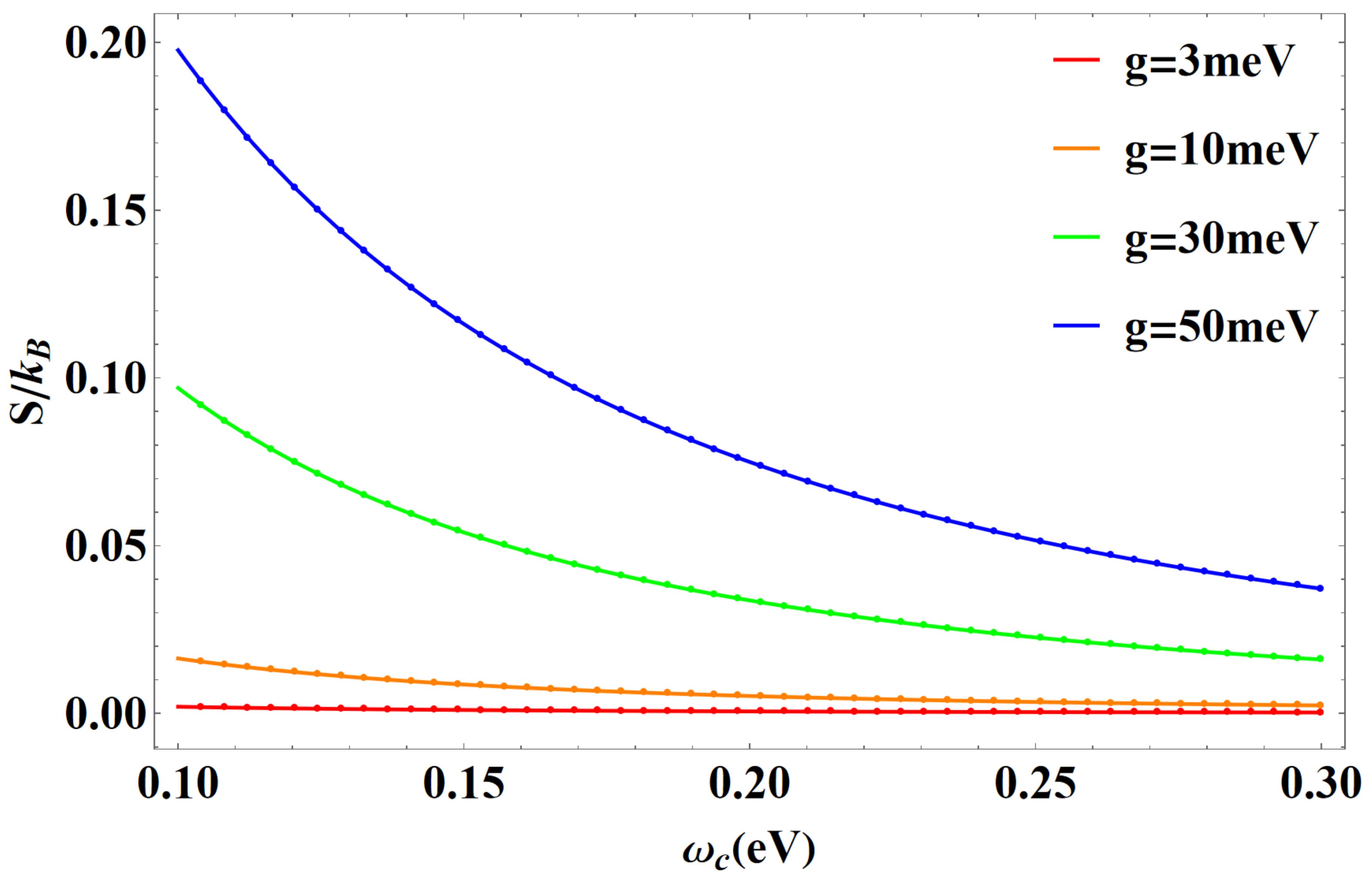}
	\caption{Entanglement entropy $S/k_B$ plotted against the bare cavity energy $\omega_{\rm c}$ with $\omega_{0}=169.1$ meV and different values of $g$.}
	\label{fig:entropy}
\end{figure}

In our ground state,
\begin{equation}
	\begin{aligned}
	\langle\hat x_1^2\rangle=\langle\hat p_2^2\rangle=\frac b2,\qquad \langle\hat x_2^2\rangle=\langle\hat p_1^2\rangle=\frac a2,\\
	\frac12\langle\hat x_1\hat p_1+\hat p_1\hat x_1\rangle=\frac12\langle\hat x_2\hat p_2+\hat p_2\hat x_2\rangle=0,
	\end{aligned}
\end{equation}
and using the simple relation $ab=1+c^2$ which can be derived from Eq. (\ref{equ:ab}) and Eq. (\ref{equ:c}), it is obvious that parameter $\sigma=\frac12\left(\sqrt{c^2+1}-1\right)=\sinh^2\left(\frac r2\right)$, where we let $c=\sinh r$. Thus the entanglement entropy can be written in a form that merely depends on $c$,
\begin{equation}
	\begin{aligned}
	S/k_B&=\sqrt{c^2+1}\ln\frac{\sqrt{c^2+1}+1}{c}+\ln\frac{c}{2}\\
	&=\cosh r\ln\left(\coth\frac r 2\right)+\ln\left(\frac{\sinh r}{2}\right).
	\end{aligned}
\end{equation}
When the coupling strength $g\rightarrow 0$, then since $\sigma\rightarrow0$, we find the entanglement entropy $S\rightarrow0$. The entanglement entropy as a function of the bare cavity energy $\omega_{\rm c}$ is presented in Fig. \ref{fig:entropy} for the phonon frequency $\omega_{0}=169.1$ meV of hBN and different coupling strength $g$. With increasing coupling strength, we get the increased entanglement entropy as expected.

\subsection{Logarithmic Negativity Parameter}

\begin{figure}[b]
	\includegraphics[width=8.6cm]{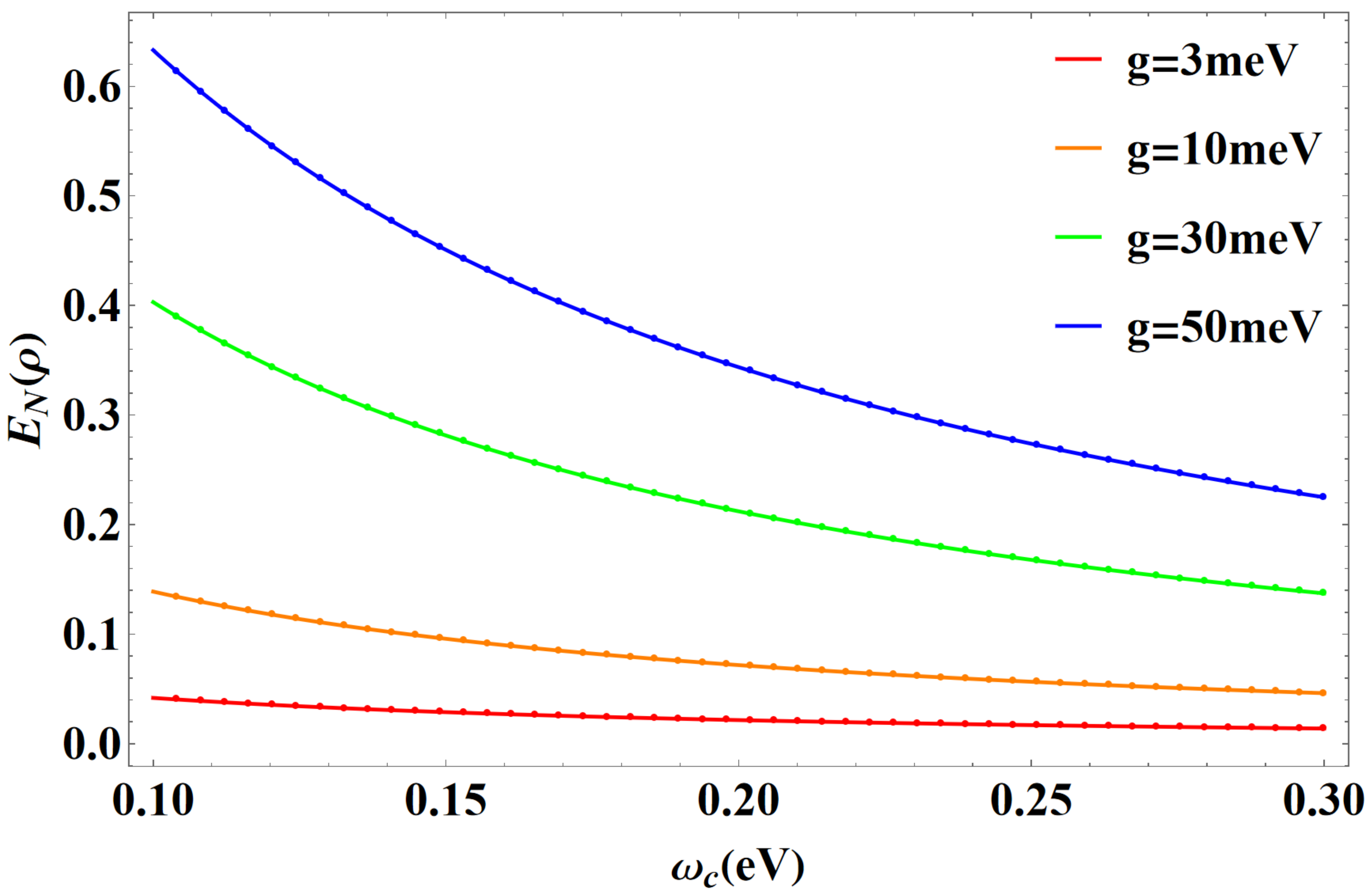}
	\caption{Logarithmic negativity $E_N(\rho)$ plotted against the bare cavity energy $\omega_{\rm c}$ with $\omega_{0}=169.1$ meV and different values of $g$.}
	\label{fig:neg}
\end{figure}

The Peres-Horodecki separability criterion to determine whether bipartite continuous variable states are separable was presented by Simon \cite{Simon2000}, and for a two-mode Gaussian state determined by variance matrix $V$, the criterion is simply an inequality obeyed by a complete set of ${\rm Sp}(2,R)\otimes{\rm Sp}(2,R)$ invariants for $V$. The elements of variance matrix $V$ are given by
\begin{equation}
	V_{ij}=\frac12{\rm tr}\left(\hat\rho\left\{\hat\xi_i,\hat\xi_j\right\}\right),
\end{equation}
where there is no displacement, and four dimensional vector $\hat\xi=(\hat x_1,\hat p_1,\hat x_2,\hat p_2)$ for our two-mode system. The variance matrix has the form $V=\left(
\begin{array}{cc}
	A & C\\
	C^T & B
\end{array}
\right)$, and the invariants are the numbers $\det A$, $\det B$, $\det C$ and $\det V$. For a pure Gaussian state, $\det V=\frac{1}{16}$, and $\det A+\det B+2\det C=\frac12$ \cite{Vidal2001}. 

The Peres-Horodecki separability criterion states that a separable density operator under the partial transpose goes necessarily into a non-negative operator, and the logarithmic negativity parameter is nothing but a quantitative version of the criterion that quantifies how much the new density operator deviates from being a non-negative matrix. 

Two sympletic eigenvalues of the covariance matrix $V$ associated with its density operator under the partial transpose can be derived \cite{Vidal2001,Agarwal2010}, 
\begin{equation}
	\tilde\nu_\pm=\sqrt{\frac{\tilde\Delta(V)-\sqrt{\tilde\Delta(V)^2-4\det V}}{2}},
\end{equation}
where $\tilde\Delta(V)=\det A+\det B-2\det C$. Equivalent to the criterion presented by Simon, $\nu_-<\frac12$ is a necessary and sufficient condition for two-mode Gaussian states to be entangled. In our ground state, the ${\rm Sp}(2,R)\otimes{\rm Sp}(2,R)$ invariants are calculated,
\begin{equation}
	\det A=\det B=\frac14(1+c^2),\qquad \det C=-\frac14 c^2,
\end{equation}
thus $\tilde\Delta(V)=\frac12+c^2$,
\begin{equation}
	\tilde\nu_{-}=\frac12(\sqrt{c^2+1}-c)=\frac12e^{-r}.
\end{equation}

Since our coefficient $c>0$, then $\tilde\nu_{-}<\frac12$, the ground wave function is entangled. When the coupling strength $g\rightarrow0$, i.e., $c\rightarrow0$, we have $\tilde\nu_-=\frac12$, then the state becomes separable. The logarithmic negativity $E_N(\rho)$ is used to measure the amount of quantum entanglement,
\begin{equation}
	E_N(\rho)=\max[0,-\log_2(2\tilde\nu_{-})].
\end{equation}
The logarithmic negativity as a function of the bare cavity energy $\omega_{\rm c}$ is presented in Fig. \ref{fig:neg} for $\omega_{0}=169.1$ meV of hBN and different coupling strength $g$. The trend of logarithmic negativity is very similar to entanglement entropy, and we can reach the same conclusion: for weak coupling, there is almost no entanglement, and we can obtain a large amount of entanglement in the ultrastrong regime of phonon-cavity interaction.

\section{Conclusions}

In the experiment, we can measure polariton energies $\omega_\pm$ from the transmission spectrum. Then by matching with Eq. (\ref{equ:en}) applying our theory model, we can determine the coupling strength of our system, and coefficient $c$ can be easily found by Eq. (\ref{equ:c}). Since the entanglement entropy and logarithmic negativity are exclusively determined by the coefficient $c$, we can make a possible quantitative measure of quantum entanglement in the ground state following these procedures. It should be noted that we do not discuss entanglement in the first excited states. These states exhibit entanglement despite not being in the ultrastrong coupling regime. 

While our theory model is centered around phonon polaritons in cavities, it can be extended to other types of polaritons. Applying our conclusion to exciton polaritons is possible \cite{KenaCohen2013}, but it is important to note that normal incidence should be assumed, i.e., the incidence angle $\theta$ must be set to zero. We should also neglect spatial dispersion, then the expression for the vacuum Rabi energy becomes $\Omega=\omega_0\sqrt{\frac{\pi\chi\omega_0}{\omega_c\epsilon_\infty}}$, where $\chi$ represents the coupling constant between photon and exciton oscillators \cite{Hopfield,Quattropani1986,Birman1991}. The relation between $g$ and $\chi$ can be established, which is given by $g=\omega_0\sqrt{\frac{\pi\chi}{\epsilon_\infty}}$. Another example of applying our theory is plasmon polaritons. As explored in a recent study \cite{plasmon2020}, the vacuum Rabi energy $\Omega$ is independent of the energy of the cavity mode $\omega_{\rm c}$. Then we can modify the interaction parameter in our theory, i.e., make $\Omega=g$. This modification leads to the same conclusion that the quantum entanglement entropy in the ground state increases with the coupling strength.

\medskip

\textbf{Acknowledgements}

We thank the support of Air Force Office of Scientific Research (Award No FA-9550-20-1-0366) and the Robert A Welch Foundation (A-1943-20210327).

\bibliography{main}

\end{document}